\documentclass[preprint,12pt]{elsarticle}

\usepackage{graphicx}
\usepackage{amssymb}
\usepackage{amsmath}
\usepackage{enumerate}
\usepackage{graphicx}
\usepackage{subfigure}
\usepackage{multirow}
\usepackage{latexsym}

\usepackage{lineno}
\begin{document}



\journal{Physica A: Statistical Mechanics and its Applications}

\begin{frontmatter}


\title{Measuring Road Network Topology Vulnerability by Ricci Curvature}



\author[csu,tju]{Lei Gao}
\author[csu]{Xingquan Liu}
\author[pku]{Yu Liu}
\author[csu-t]{Pu Wang}
\author[csu]{Min Deng}
\author[sju]{Qing Zhu}
\author{Haifeng Li\corref{cor1}\fnref{csu,Henan}}
\ead{lihaifeng@csu.edu.cn}
\cortext[cor1]{Corresponding author}

\address[csu]{School of Geosciences and Info-Physics, Central South University, South Lushan Road, Changsha, China}
\address[tju]{The Key Laboratory of Road and Traffic Engineering, Ministry of Education, Tongji University, Shanghai, China}
\address[pku]{School of Earth and Space Sciences, Peking University, Beijing, China}
\address[csu-t]{School of Traffic Transportation Engineering, Central South University, Changsha, China}
\address[sju]{Faculty of Geosciences and Environmental Engineering, Southwest Jiaotong University, Chengdu, China}
\address[Henan]{Henan Laboratory of Spatial Information Application on Ecological Environment Protection, Zhengzhou 450000, China}


\begin{abstract}
Describing the basic properties of road network systems, such as their robustness, vulnerability, and reliability, has been a very important research topic in the field of urban transportation. Current research mainly uses several statistical indicators of complex networks to analyze the road network systems. However, these methods are essentially node-based. These node-based methods are more concerned with the number of connections between nodes, and lack of consideration for interactions. So, this leads to the well-known node paradox problem, and their ability of characterizing the local and intrinsic properties of a network is weak. From the perspective of network intrinsic geometry, this paper proposes a method for measuring road network vulnerability using a discrete Ricci curvature, which can identify the key sections of a road network and indicate its fragile elements. The results show that our method performs better than complex network statistics on measuring the vulnerability of a road network. Additionally, it can characterize the evolution of the road network vulnerability among different periods of time in the same city through our method. Finally, we compare our method with the previous method of centrality and show the different between them. This article provides a new perspective on a geometry to analyze the vulnerability of a road network and describes the inherent nature of the vulnerability of a road system from a new perspective. It also contributes to enriching the analytical methods of complex road networks.
\end{abstract}


\end{frontmatter}


\section{Introduction}
\label{T:1}

Reliable road network and other related infrastructure provide continuous transportation services for modern life. If the network is interrupted, it will not only impede residents’ daily travel but also lead to huge repair costs. For these reasons, ensuring the continuous operation of road networks is an important research topic. To solve problems posed by interruptions, a key idea is to measure vulnerability \cite{berdica2002an}, which measures the road network’s ability to function properly in the event of attacks and failures \cite{reggiani2015transport}. Vulnerability denotes the potential factors that contribute to degraded road network and the maximum serviceable level that can be maintained when the network is damaged. In essence, the topology determines the performance of a network. Therefore, the central issue in the study of vulnerability is to investigate the relationship between the road network topology and road network vulnerability \cite{agarwal2001vulnerability,Este2003Network,lopez2017vulnerability,murray2007critical}.

Road network vulnerability is usually represented by removing certain elements of a network and comparing the network connectivity before and after the removal. The removal is usually based on topological indicators of the road network \cite{mattsson2015vulnerability}. Road network topology surveys based on statistical indicators of complex networks have been a heated topic of debate in recent years \cite{jenelius2006importance,jenelius2015road,zheng2018understanding}. These indicators are usually node-based methods\cite{weber2017characterizing} such as degree\cite{chen2012identifying}, centrality\cite{gao2013aphysica}, community\cite{duan2014robustness}, clustering coefficients\cite{chen2013Identifying}. Specifically, one prominent feature of the node-based approach is that it focuses on the quantitative characteristics of nodes \cite{weber2017characterizing}. For example, the node degree describes the number of direct connections between nodes, and the betweenness centrality describes the number of times a node (edge) is passed by the shortest path.

The node-based method answers questions about the quantitative aspect of connections, such as `whether there is a connection between nodes A and B' or `the number of nodes connected to node A'. However, in terms of describing the interaction of connection, such as `how node A connects with node B' or `how to easily transfer from node A to node B', is limited. The essence of the vulnerability survey is to maintain the transmission capacity. That is, the interaction information between nodes is used as an indicator of vulnerability that the node-based method cannot provide. In addition, simply adopting the node-based method may lead to substantial differences between the model and the real world situation \cite{weber2017characterizing}. Studies have shown that a statistically significant commonality network may exhibit completely different network properties \cite{milo2004superfamilies}.

Therefore, the latest research of network geometry, Ricci curvature \cite{ollivier2007ricci,ollivier2009ricci}, is adopted in this paper to capture the topological characteristics of road networks. Unlike the node-based method, the Ricci curvature is an edge-based approach \cite{weber2017characterizing}. It measures the transmission status between nodes through the optimal transmission measure and captures both quantitative and interactive information. The contributions of this paper are as follows:

\begin{enumerate}[(1)]
    \item From the perspective of Riemannian geometry, we examine the topology of the road network, especially the state of topology transmission, through the Ricci curvature method. This method helps us understand the relationship between road network topology and vulnerability.
    \item The large amount of negative curvature in the road network inspired us to examine the role of edges with a negative curvature in the road network. Moreover, a negative curvature implies that the vulnerability of road network could be shown in the research by simulated experiments. In addition, the vulnerability characteristics of different types of road networks can be found through the detailed investigation of the road networks among six different cities. We concurrently reviewed the vulnerability characteristics during the process of road network evolution. 
    \item We compared different methods that study road network vulnerability and demonstrated the relationship between our model and the betweenness centrality model.
\end{enumerate}

The structure of this paper is as follows: in the section of the related works, previous research on road network vulnerability is briefly reviewed. The method of using Ricci curvature to measure road network vulnerability will be described in detail in the section of the methods. In the result and analysis section, three specific experiments designed in this paper will be shown. Further work and conclusion will be put forward in the last part, followed by the discussion.

\section{Related Works}
\label{T:2}
The existing indexes of road network topology can generally be divided into global and local types\cite{lu2016vital}. In studying the topology vulnerability of a road network, the global index focuses on the global properties of the road network, such as the scale-free characteristics, the small world characteristics, etc., while the local index describes the key nodes and edges in the network to indicate elements that are easily interrupted, such as all kinds of centrality indexes.

For global indicators, the degree distribution is the basic indicator to describe network properties. For example, a regular grid or random network mostly has a homogeneous degree distribution, while a scale-free network and a small-world network mostly have a heterogeneous degree distribution \cite{sole2004information}. The key to scale-free attributes of a network is whether the degree distribution has a power law distribution \cite{barabasi1999emergence,jacob2017robustness}. The power law distribution indicates that some nodes in the network have a higher number of connections, while most nodes have low connections. Studies on network vulnerability have shown that this type of network illustrates a high robustness during random attacks, but its performance degrades rapidly during purposeful attacks\cite{AlbertAttack2000,angeloudis2006large,derrible2010the}.

However, networks with similar statistics may show completely different properties\cite{milo2004superfamilies}, meaning that multiple networks with the same statistical significance cannot be used as a basis for studying network vulnerability. Furthermore, the degree distribution focuses on the influence of node quantitative connections and ignores the interaction between nodes. In the measurement of road network vulnerability, current research has shown that a group of nodes with a high correlation can provide more abundant structural information than a single node \cite{duan2014robustness,jiang2008self-organized}. That is to say, we should focus on the interaction information posed by a set of nodes rather than a single node.

At the local level, one of the most popular vulnerability indicators is the betweenness centrality, which expresses the number of times the shortest path passes by the edge (node) in the network. Removing the section with a higher betweenness centrality value will affect the network performance \cite{holme2002attack,demsar2008identifying,matisziw2009exploring,gao2013understanding}. However, the inherent defect of betweenness centrality is that it is based on the hypothesis of the shortest path \cite{Chehreghani2007Discriminative}. Specifically, first, residents do not just choose the shortest route, relying on high-tech tools for daily travel, they may instead choose the route minimizing the total cost of transportation, such as the cost of time, distance, and tolls. Furthermore, road networks that have the same betweenness centrality value may have quite different performance values \cite{mattsson2015vulnerability}. Finally, betweenness centrality does not consider the interaction characteristics between nodes (edges).

Network interruption or degraded performance occurs when the traffic flow in some areas or edges in the network exceeds its own capacity constraints and is incapable of spreading to other parts of the network \cite{reggiani2015transport,huang2018a,cats2018beyond}. In other words, measuring the relationship between nodes, as well as the connection information in the local area, is the key to examining the network’s performance. However, whether it is global or local indicators, the node paradox\cite{ballouz2016egad,weber2017characterizing} of focusing on the independent information of the node and thus neglecting the relationship cannot be avoided in the traditional manner. Recently, some studies have considered the network topology as based on the interaction of network nodes (edges) \cite{Coates2000Network,coates2002maximum}. The Ricci curvature in the network geometry is one of these methods, which advantageously captures the relationship between nodes in the local neighborhood. Therefore, inspired by the previous research from the Internet \cite{ni2015ricci}, financial networks \cite{sandhu2016ricci}, and biological networks \cite{sandhu2015graph,tannenbaum2015graph}, in this paper we measure road network vulnerability based on the Ricci curvature.


\section{Methods}
\label{T:3}
\subsection{Overview}
\label{ST:3:1}

The vulnerability of the road network topology measures the performance of the components in the road network through topology information. In the past, reviews of topological information focused on independent information for intersections and roads. However, the transmission function of the road is affected by a set of nodes or edges, that is its neighbors, and a vulnerability is caused by a mismatch in transmission conditions between neighborhoods. To this end, we divide the road network topology vulnerability research into three steps, as shown in \textbf{\textit{Figure}} \ref{Figure 1}.

\begin{figure}
\label{Figure 1}
\centering\includegraphics[width=0.9\linewidth]{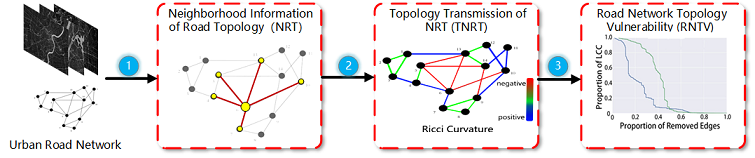}
\caption{Framework of topology vulnerability measurement method based on Ricci curvature}
\label{Figure 1}
\end{figure}

\begin{enumerate}[(1)]
    \item Neighborhood Information of Road Topology (NRT)：Since road transmission is interactive information, the method of independently examining intersections and roads will eliminate interactive effects. To avoid this problem, we captured road neighborhood information for further examination. Detailed NRT methods are shown in Section \ref{ST:3:2}.
    \item Topology Transmission of NRT (TNRT): A comprehensive measurement of road topology based on the Ricci curvature. We used the Ricci curvature to measure neighborhood information for each road and used it as a basis for measuring vulnerability. See Section \ref{ST:3:3} for more details.
    \item Road Network Topology Vulnerability (RNTV): empolymenting the results in Section \ref{ST:3:3}, we investigated the Largest Connected Component (LCC) of road networks when facing road events through different simulation attacks. More information about RNTV is described in Section \ref{ST:3:4}.
\end{enumerate}

\subsection{Neighbors’ Information of Road Topology (NRT)} 
\label{ST:3:2}

Neighbors’ Information of Road Topology (NRT) is the use of probability measures to describe the topology information of road neighborhoods. The edges in the road network describe the topology information of two adjacent roads. The traffic flow of an adjacent road in the road network is not only related to the topology information of the current road but it is also related to all the neighbor’s topologies constituting the road, such as the intersections shown in \textbf{\textit{Figure}} \ref{Figure 2}. The edge structure in the traditional road network model only contains the topology of the current road. In addition, the traffic flow of the road network itself has a high degree of uncertainty and dynamics. Therefore, we describe the NRT with a probability measure, and then the traffic flow between two adjacent roads can be easily characterized by the distance between the probabilities. 

We employment a probability distribution to describe NRT. Specifically, using $(X,d)$ as a metric space, there is a probability measure $m_X (\cdot)$ for each point $x\in{}X$.Similarly, we have $m_Y (\cdot)$ for $y\in{}Y$. Then, we calculate the measure by the \textbf{\textit{Eq 1}}, where the $d_X$ refers to the node's degree.

\begin{equation}\label{Eq1}
m_X(x) =  \begin{cases} \frac{1}{d_X},& \text{if $x \sim X$ ;} \\ 0,& otherwise. \end{cases}
\end{equation}

\subsection{Topology Transmission of NRT (TNRT)} 
\label{ST:3:3}

The Topology Transmission of NRT (TNRT) describes traffic flow under the neighborhood topology of connected roads. When we express the road neighborhood topology information by the probability measure, we can use the distance between the probabilities to describe the traffic flow between two adjacent roads. as shown in \textbf{\textit{Figure}} \ref{Figure 2}. The distance between probabilities can be characterized by a Kullback-Leibler divergence (KL divergence) and Wasserstein distance. Since the Wasserstein distance has better geometric and mathematical properties relative to the KL divergence, this paper uses the Wasserstein distance to calculate the distance between topological neighborhoods, see the \textbf{\textit{Eq 2}}.

\begin{figure}
\label{Figure 2}
\centering\includegraphics[width=0.9\linewidth]{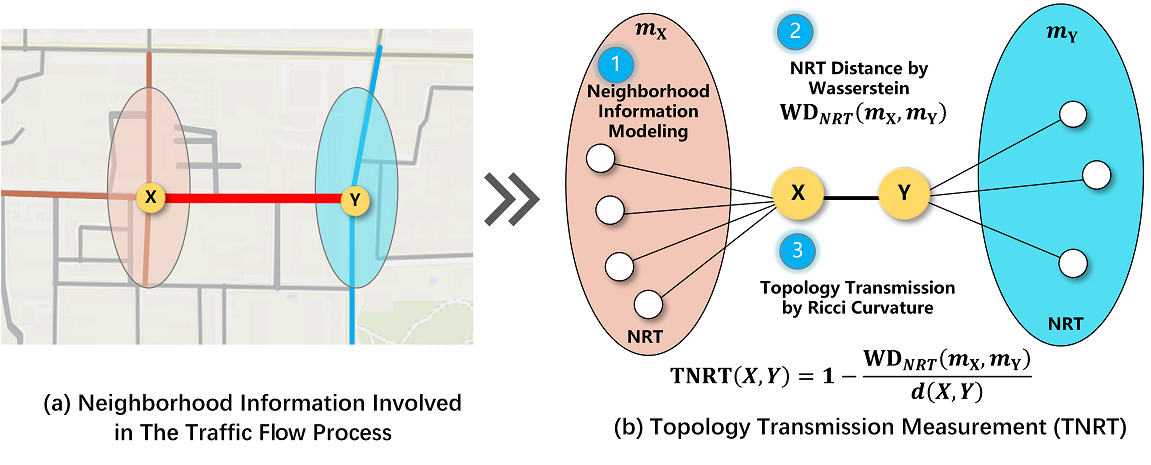}
\caption{Topology transmission of NRT measurement (TNRT)}
\label{Figure 2}
\end{figure}

\begin{equation}\label{Eq2}
    WD_{NRT}(m_{X},m_{Y}) = \inf_{\xi} \int d(X,Y)d_{\xi}(X,Y)
\end{equation}

The definition of the Wasserstein distance is similar to solving a transmission planning problem.We informally introduce them bellow and detailed information can be found in\cite{RN832,RN833}. Specifically, we can use a function $\xi(X,Y)$ to describe a transport plan $\xi$, the function $\xi(X,Y)$ gives the amount of mass to move from X to Y. In order to solve this plan, the following properties are required:
\begin{equation}\label{Eq3}
    \int \xi(X,Y)dx = m(X)
\end{equation}
\begin{equation}\label{Eq4}
    \int \xi(X,Y)dy = m(Y)
\end{equation}

That is, the mass transmitted from x must be equal to $m(X)dx$ and the total mass moved into y must be equal to $m(Y)dy$. In other words, $\xi$ is a joint probability distribution with marginals X and Y. Therefore, the total mass transported from X to Y is $\xi(X,Y)dxdy$. Then we specify the cost d(X,Y) from x to y, and the cost moving is $d(X,Y)\xi(X,Y)dxdy$. Finally, the total cost of a transport plan $\xi$ is:

\begin{equation}\label{Eq5}
\int\int d(X,Y)\xi dxdy = \int d(X,Y)d_{\xi}(X,Y)
\end{equation}

The goal is to minimize transmission costs, so we get the \textbf{\textit{Eq2}}.

Then, we measure TNRT based on the Ollivier’s coarse Ricci curvature. In general, curvature describes the degree to which a curve (surface) deviates from a straight line (flat plane). The most common curvature is Gaussian curvature and section curvature, and Ricci curvature can be constructed by section curvature. We briefly introduce them and detailed information can be found in\cite{ollivier2007ricci,ollivier2009ricci}. We consider a point $X$ on a surface $M$ and gives the tangent vector $V$ of the point, then $V$ intersects at point $Y$ on $M$. Take another tangent vector $V\_{X}$ at point $x$, and then move $V\_{X}$ along $V$ to the point $Y$, marking it as $V\_{y}$. So, we can get the endpoints of $V\_{X}$ and $V\_{y}$ as $x^{'}$ and $y^{'}$, seen in \textbf{\textit{Figure}} \ref{Figure 3}. Then the difference of distance between $x^{'}$ and $y^{'}$ to $|V|$ can be used as a basis for measuring section curvature.

\begin{figure}
\label{Figure 3}
\centering\includegraphics[width=0.9\linewidth]{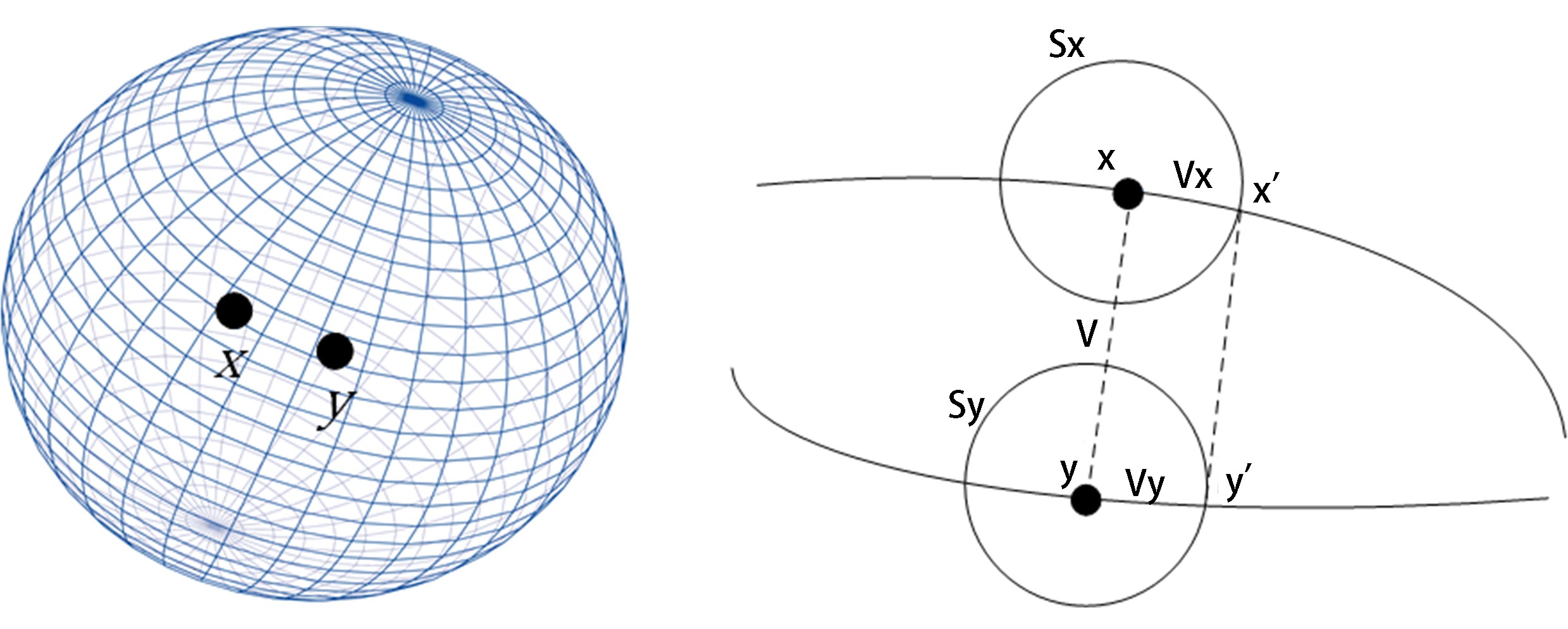}
\caption{Section curvature and the intuition of Ollivier’s Ricci curvature}
\label{Figure 3}
\end{figure}

According to the \textbf{\textit{Figure}} \ref{Figure 3}, we can find that the section curvature depends on two tangent vector named $V$ and $V_{X}$. In fact, we can make an infinite number of tangent vectors in different directions at point $x$, so different section curvatures may be formed in different directions. The Ricci curvature is formed by the average section curvatures and depends on one direction. If we think of a small geodesic ball $S_{X}$ with the radius of $V$ in $X$ on $M$, the Ricci curvature measures that whether the distance between the two small balls named $S_{X}$ and $S{y}$ is less or greater than the distance between the small balls’ centers. The distance between the balls can be measured using the Wasserstein distance. TNRT is not only related to the transfer distance, but is also related to the transfer path. See the \textbf{\textit{Eq6}}.

\begin{equation}\label{Eq6}
TNRT(X,Y)=1- \frac{ WD_{NRT}(m_{X},m_{Y})}{d(X,Y)}
\end{equation}

where $WD_{NRT}(m_{X},m_{Y}) $ refers to transmission distance between node $X$ and node $Y$.
Assuming that $d(X,Y)=1$, that is, when the topology transfer is performed, the transmission path is not considered, the \textbf{\textit{Eq6}} can be changed to:

\begin{equation}\label{Eq7}
    TNRT(X,Y)=1-  WD_{NRT}(m_{X},m_{Y})
\end{equation}

\subsection{Road Network Topology Vulnerability (RNTV)}
\label{ST:3:4}
In this paper, the connectivity performance is used as a measure to quantify the vulnerability when the network is attacked. Specifically, when some of the edges are disrupted, if the network connectivity drops rapidly, the network is then vulnerable. We use the Largest Connected Component (LCC) to express the network connectivity. If a network is a connection graph, then LCC is the number of network nodes. When some nodes are removed, the network will lose its global connectivity and be divided into $N$ subgraphs. Therefore, the LCC expresses the maximum number of nodes connected, see \textbf{\textit{Eq8}}.

\begin{equation}\label{Eq8}
    TVR =  \frac{LCC_{RG^{'}}}{E_{RG}}
\end{equation}
Where: $E_{RG}$ represents the number of nodes in the original road network. $LCC_{RG^{'}}$ indicates the number of nodes in the largest connected subgraph after the attack.

Road network vulnerability measures the road network’s performance during attacks. We designed different attack patterns to simulate the reaction of the urban road network during these attacks. Road disruptions can be normally divided into two categories \cite{jenelius2015road}. One is natural disasters such as floods, tsunamis, earthquakes, and debris flows, and the other is external events, which are generally unpredictable and random and are usually caused by man-made events that are purposeful and occur on that specific path. Therefore, we designed two experimental models to test the road network: random attacks and intentional attacks. The former is carried out by a random selection of edges in the road network, while the latter is carried out by a given attack order.

\section{Result and Analysis}
\label{T:4}
Different cities have different road network planning and construction principles as well as topology. We introduced our data in Section \ref{ST:4:1}. We endeavored to understand the vulnerability expressions under different road network topologies. We show the results of these experiments in Section \ref{ST:4:2}. In addition, in the same city, the road network would change with the development of the city. We looked to obtain the expression of vulnerability during the evolution of the road network, and the results are shown in Section \ref{ST:4:3}. Finally, we compared the traditional methods and the methods used in this article, and the details are shown in Section \ref{ST:4:2}.

\subsection{Datasets and Road Network Modeling}
\label{ST:4:1} 

\subsubsection{Datasets}

In this chapter, we will introduce the road network data set used and describe the calculation of the Ricci curvature in detail.
The \textbf{\textit{Table}} \ref{description} detail the data used in this article. Networks often have two main types of spatial effects on traffic flows, centrifugal and centripetal \cite{rodrigue2016book} as shown in \textbf{\textit{Figure}} \ref{Figure 4}. Centrifugal networks usually have a grid-like type that has the effect on spreading traffic. Conversely, centripetal networks have a star-like type that has a pooling effect. To fully investigate the properties of different types, we have chosen six different urban road networks \cite{chen1993}. The following list details each road network:

\begin{table}[]
\centering
\caption{Data description. D is the diameter of the network, L is the average shortest path of the network, C is the average clustering coefficient of the network.}
\label{description}
\resizebox{\textwidth}{!}{%
\begin{tabular}{|c|c|c|c|c|c|c|c|c|c|}
\hline
\multicolumn{2}{|c|}{Description} & Datasets & Nodes & Edges & Max Degree & Avg Degree & D & L & C \\ \hline
\multirow{3}{*}{CENTRIPETAL} & Grid-Radial-Circle & Bj2008 & 3124 & 5044 & 6 & 3.23 & 70 & 27.69 & 0.018 \\ \cline{2-10} 
 & Grid-Sector & Shenyang & 4085 & 6537 & 6 & 3.2 & 57 & 25.24 & 0.019 \\ \cline{2-10} 
 & Grid-Star & Changchun & 6376 & 10362 & 7 & 3.25 & 58 & 26.86 & 0.02 \\ \hline
\multirow{3}{*}{CENTRIFUGAL} & \multirow{3}{*}{Grid-based} 
 & Luoyang & 2742 & 4135 & 5 & 3.02 & 53 & 24.45 & 0.02 \\ \cline{3-10} 
 &  & Xi'an & 8519 & 12826 & 7 & 3.01 & 73 & 33.49 & 0.018 \\ \cline{3-10} 
 &  & NYC-Manhattan & 6655 & 11185 & 6 & 3.36 & 125 & 48.75 & 0.018 \\ \hline 
\end{tabular}%
}
\end{table}

\begin{itemize}
\item Grid-Radial-Circle. This type of network is usually organized in a grid form, while the radiation form is used among the regions. There is also a loop-linked layout (e.g., Bj) for this network type. We also extracted historical data from each year of Bj to show the changing characters during different periods as shown in \textbf{\textit{Table}} \ref{Bj-data}.
\item Grid-Sector combination. The base type is grid, and the main road is connected by sectors (e.g., Shenyang).
\item Grid-Star. The urban trunk road system is composed of several star network combinations, and the branch is grid type (e.g., Changchun).
\item Grid based. An urban road system is usually composed of a grid (e.g., Xi'an, Luoyang and NYC-Manhattan), yet the scales of Luoyang and Xi'an are different. To compare the differences between Chinese and foreign cities, we introduce the network of NYC-Manhattan for evaluation.
\end{itemize}

\begin{table}[]
\centering
\caption{Bj-data description for each year. D is the diameter of the network, L is the average shortest path of the network, C is the average clustering coefficient of the network.}
\label{Bj-data}
\begin{tabular}{|c|c|c|c|c|c|c|c|}
\hline
Datasets & Nodes & Edges & Max Degree & Avg Degree & D & L & C \\ \hline
Bj1969 & 290 & 428 & 5 & 2.95 & 25 & 10.45 & 0.005 \\ \hline
Bj1978 & 871 & 1332 & 6 & 3.06 & 41 & 16.55 & 0.01 \\ \hline
Bj1990 & 1280 & 2017 & 6 & 3.15 & 46 & 19.03 & 0.01 \\ \hline
Bj2000 & 2267 & 3561 & 6 & 3.14 & 60 & 24.13 & 0.015 \\ \hline
Bj2008 & 3124 & 5044 & 6 & 3.23 & 70 & 27.69 & 0.018 \\ \hline
\end{tabular}
\end{table}

In the tables, D is the diameter of the network, meaning that the maximum number of edges between any two nodes in the network can be connected. L is the average path length of the network, which means that the average number of edges between any two nodes in the network can be connected. C is the average clustering coefficient of the network. In this paper, we assume that the network is undirected and unweighted. We use the average clustering coefficient of each node in the network to calculate C, indicating the tightness of the node. The clustering coefficient $c_{i}$ of node $i$ is
\begin{equation}\label{Eq9}
    c_{i} = \frac{2*m_{i}}{k_{i}(k_{i}-1)}
\end{equation}
Where, $k_{i}$ is the degree of node $i$, $m_{i}$ is the number of links between the nodes connected with $i$. It represents the ratio of the actual connections among neighbors of the node $i$ to the possible connections, reflecting the degree of tightness between nodes.

\begin{figure}
\label{Figure 4}
\centering\includegraphics[width=0.8\linewidth]{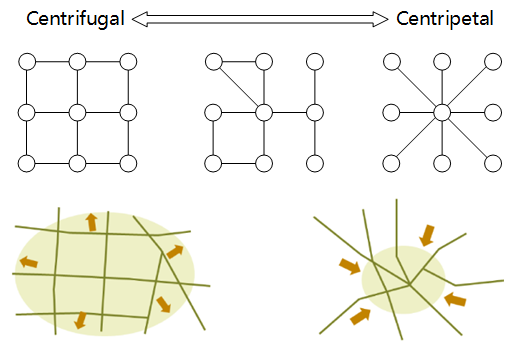}
\caption{Centrifugal and centripetal road networks\cite{rodrigue2016book}}
\label{Figure 4}
\end{figure}

\subsubsection{Road network Construction}
 
There are various types of urban public transport network modeling, and different methods are suitable for different applications \cite{vonferber2009public}. For example, the L-space model uses nodes to represent road intersections (metro, bus stops), and edges of the road between two intersections (subway line). A ring is not allowed. This modeling approach shows the actual topology of the transport system and investigates its topological properties; the nodes of B-space represent the road (metro, bus line) while the edge (metro, bus line) models the exact link between the edges of the road. Note that the edges are virtual in this mode. This modeling method denotes number of links on each road in the road system and the exact relationship between roads. Under this modeling method, the power law distribution of the road network can be seen. P-space is the approach used to build a bipartite graph. In this study, we use L-space to explore the topology of the road network with the intersection as the nodes and the road as the edges as shown in \textbf{\textit{Figure}} \ref{Figure 5}. We can see some basic statistics about our network from \textbf{\textit{Table}} \ref{description} and \textbf{\textit{Table}} \ref{Bj-data}. It can be seen that in our construction mode, two different types of networks show consistent statistical results. Therefore, the characteristics of the network cannot be detected in detail in this mode.

\begin{figure}
\label{Figure 5}
\centering\includegraphics[width=0.9\linewidth]{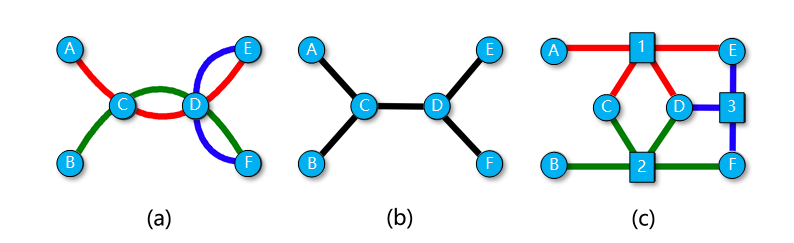}
\caption{Road network construction. (a)original road network (b) L-space model (c) B-space model}
\label{Figure 5}
\end{figure}

\subsection{Vulnerability Analysis of Different Types of Road Networks} 
\label{ST:4:2} 

In this section, we will discuss the numerical distribution of the Ricci curvature in Section 4.2.1 and point out the performances of indicators used in TNRT measurements in this study. This will be followed by an examination of the vulnerability of different types of road networks in Section 4.2.2.

\subsubsection{TNRT Analysis}
\label{ST:4:2:1}
The topology transmission measurement based on Ricci curvature measures the road network’s topology. In \textbf{\textit{Figure}} \ref{Figure 6}, we use histograms to show the numerical distribution of Ricci curvature in different types of road networks.

There are different topology transmission characteristics between the two network types. Specifically, the distribution of positive values for the centripetal road network is less, and the negative values’ distribution is more. For example, the peak values of Bj2008 in the centripetal network are around -0.5, and by a rough estimation the total number of negative edges is three times of the total number of positive edges. The peak values of Changchun and Shenyang are all near -0.1. By contrast, centrifugal networks have an obvious aggregation in positive distribution, such as in Xi'an, Luoyang, and NYC-Manhattan. Specifically, NYC-Manhattan has a high level of 0 value distributions, reaching approximately 50 \% for the whole network.

In addition, although Luoyang, Xi'an and NYC-Manhattan have the same type of roads, they show different topology transmission characteristics. Specifically, Luoyang has essentially the same layout as Xi'an, while NYC-Manhattan is different from them. This may be due to Luoyang and Xi'an also having ring and star road networks while NYC-Manhattan only has a grid. At the same time, we also show that the impact of network size on the topology transmission is small. Although the Luoyang and Xi'an networks have different network sizes, there is no obvious difference in TNRT distribution.

\begin{figure}
\label{Figure 6}
\centering\includegraphics[width=0.9\linewidth]{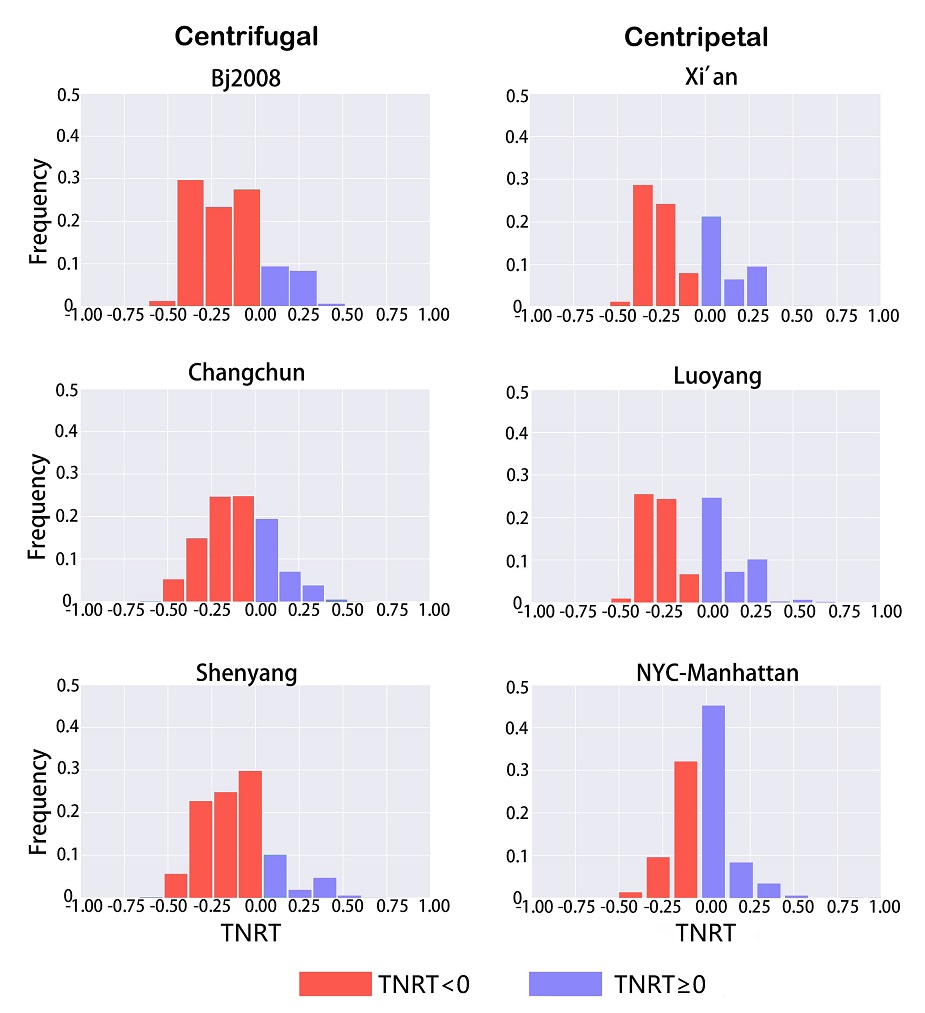}
\caption{Distribution of TNRT values of road network in different cities}
\label{Figure 6}
\end{figure}

Furthermore, to obtain the actual representation of Ricci curvature values on the road network, especially for the meaning of negative curvature, we visualize the curvature distribution of the road network in six cities as shown in \textbf{\textit{Figure}} \ref{Figure 7}.

The properties of the road can be reflected by the TNRT measurement. Specifically, the red color represents a negative curvature and blue represents a positive curvature in all urban road networks. We can roughly indicate that the positive curvature (blue) identifies both `branch paths' or `separate paths' and negative values (red) mean `main paths' which connect two regions.

\begin{figure}
\label{Figure 7}
\centering\includegraphics[width=0.9\linewidth]{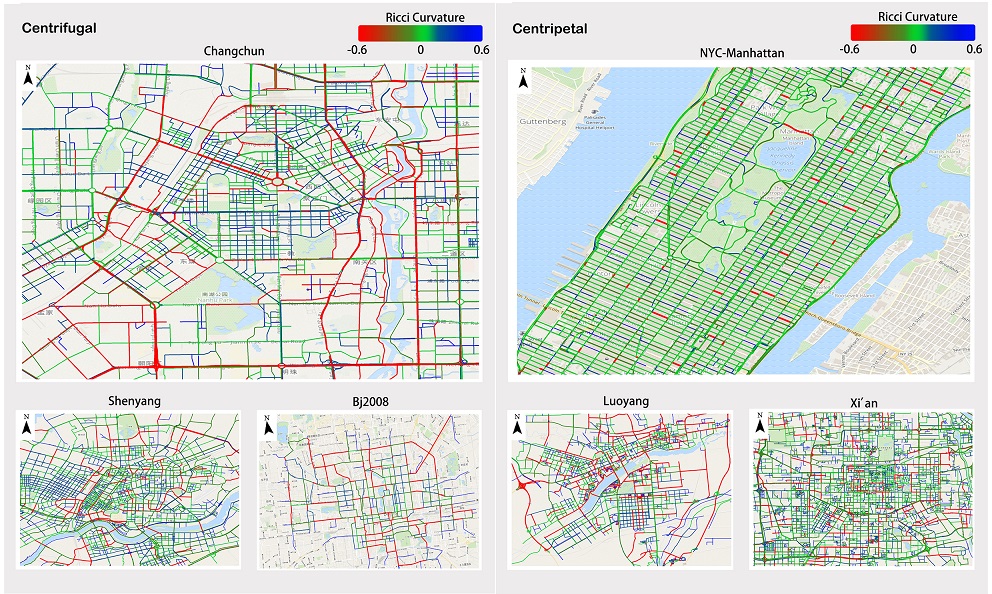}
\caption{Spatial distribution of TNRT values}
\label{Figure 7}
\end{figure}


\subsubsection{TVR Analysis}
\label{ST:4:2:2}
As seen from the TNRT distribution in the section \ref{ST:4:2:1}, the negative curvature distribution is prominent in all road networks, motivating us to investigate the meaning of the negative curvature. In this section we employment the experimental scheme of section \ref{ST:3:4} (random and targeted attacks) to explore the TVR of the road network, especially the function of the negative curvature. The result is shown in \textbf{\textit{Figure}} \ref{Figure 8}.

Blue lines are shown as targeted attacks starting at the most negative curvature and then increasing in the order of Ricci curvature. Green lines are random attacks. Compared with random simulation attacks, we find that intentional attacks are more harmful to the road network, regardless of whether it is a centrifugal or centripetal network. Specifically, the declining level of the LCC in the road network shows that the rate of decline in intentional attacks is faster than that of the random attacks. In general, when 20\% of the road network is attacked purposefully, the entire network can only maintain 20\% of its normal operating capacity. However, at the same attack scale, the whole network can maintain 80\% normal operating capacity during random attacks. The results show that all road networks illustrate vulnerabilities under random attacks and a small range of attacks will affect the performance of the entire network, suggesting that strengthening the protection of roads containing a negative curvature and awareness of deliberate attacks are needed.

\begin{figure}
\label{Figure 8}
\centering\includegraphics[width=0.9\linewidth]{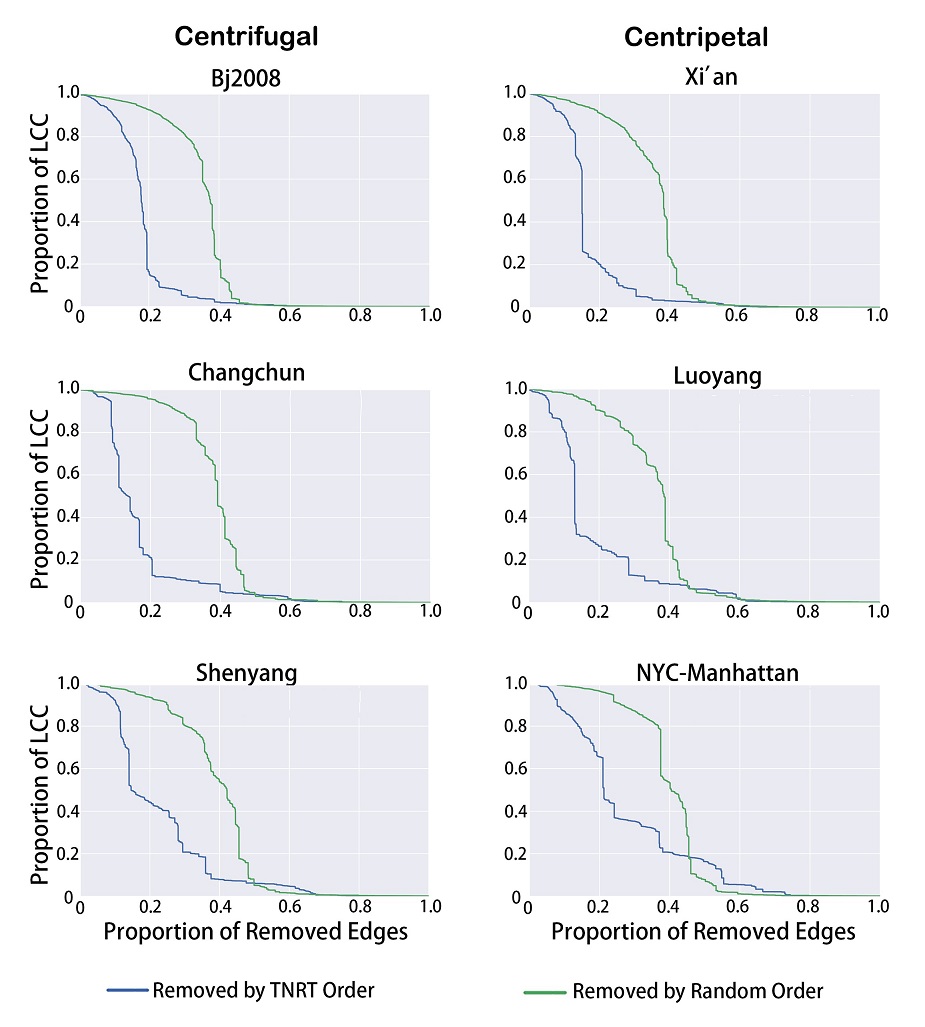}
\caption{TVR performance of road networks under different attack modes}
\label{Figure 8}
\end{figure}

When the attack scale expands to 50\%, the entire road network cannot operate effectively whether the attack is random purposeful. Unlike other studies, such as for Internet topology, when the scale of attack amplifies to 50\%, the network can maintain nearly 60\% of the operating capacity in the random attack mode, while the network fails in the purposeful attack mode (see \cite{ni2015ricci}). These results show that road networks are different from other network topologies.

In addition, the results also reflect that centrifugal and centripetal networks are different when faced with attacks, and the vulnerability of the centrifugal network is slightly lower than that of the concentric network. Specifically, for a centrifugal network composed of many grids such as NYC-Manhattan, when the attacks’ range was extended to 30\%, the region’s road network can keep approximately 40\% of the network running normally. This indicates that the NYC-Manhattan area's road network is more robust than those of other cities during the intentional attack mode. 

The network of the NYC-Manhattan area during the random attack mode also shows a high robustness. When the attack size reaches 40\%, the other cities only operated at 40\% capacity, whereas the road network in the NYC-Manhattan area can maintain nearly 70\% of the road network. Therefore, whether it is an intentional attack or a random attack, the road network in the NYC-Manhattan area is superior to those in other cities by showing a high degree of robustness. Because of the inclusion of other concentric networks, the vulnerabilities of the Luoyang, Xi'an grid type networks are higher than that of NYC-Manhattan.

\subsection{Evolutionary Characteristics of Vulnerability}
\label{ST:4:3}
Over time, a road network gradually expands and evolves with a city’s expansion. As determined by the goals of economic construction and urban development, the evolution of a road network has its own specific characteristics. Understanding the characteristics of these road evolution processes, especially the vulnerability, is helpful in detecting defects in road planning and assists in building a more resilient road network. In this section, we will explore a road network’s vulnerability during its evolution by adopting the road network of Beijing at different stages of construction and conducting experiments according to the research framework of road network vulnerability. The results are as follows:

We first observe the distribution of road network curvature. From these observations, as shown in \textbf{\textit{Figure}} \ref{Figure 9}. we find that the negative distribution of the TNRT value of Beijing road network is constantly prominent, and the peak of distribution is concentrated on negative values. Except for the obvious changes from Bj1969 to Bj1978, the variation in other periods is relatively small. Specifically, most negative values of the Bj1969 network are between -0.2 and 0 and fewer are distributed in the $<$-0.2 interval with a frequency below 0.1. However, after Bj1978, the peak value of negative values appeared at approximately -0.2 for each year, and gradually decreased with -0.2 as the center. Bj1969 is sparse in the positive distribution. After Bj1978, the distribution of positive values increased gradually.

\begin{figure}
\label{Figure 9}
\centering\includegraphics[width=0.75\linewidth]{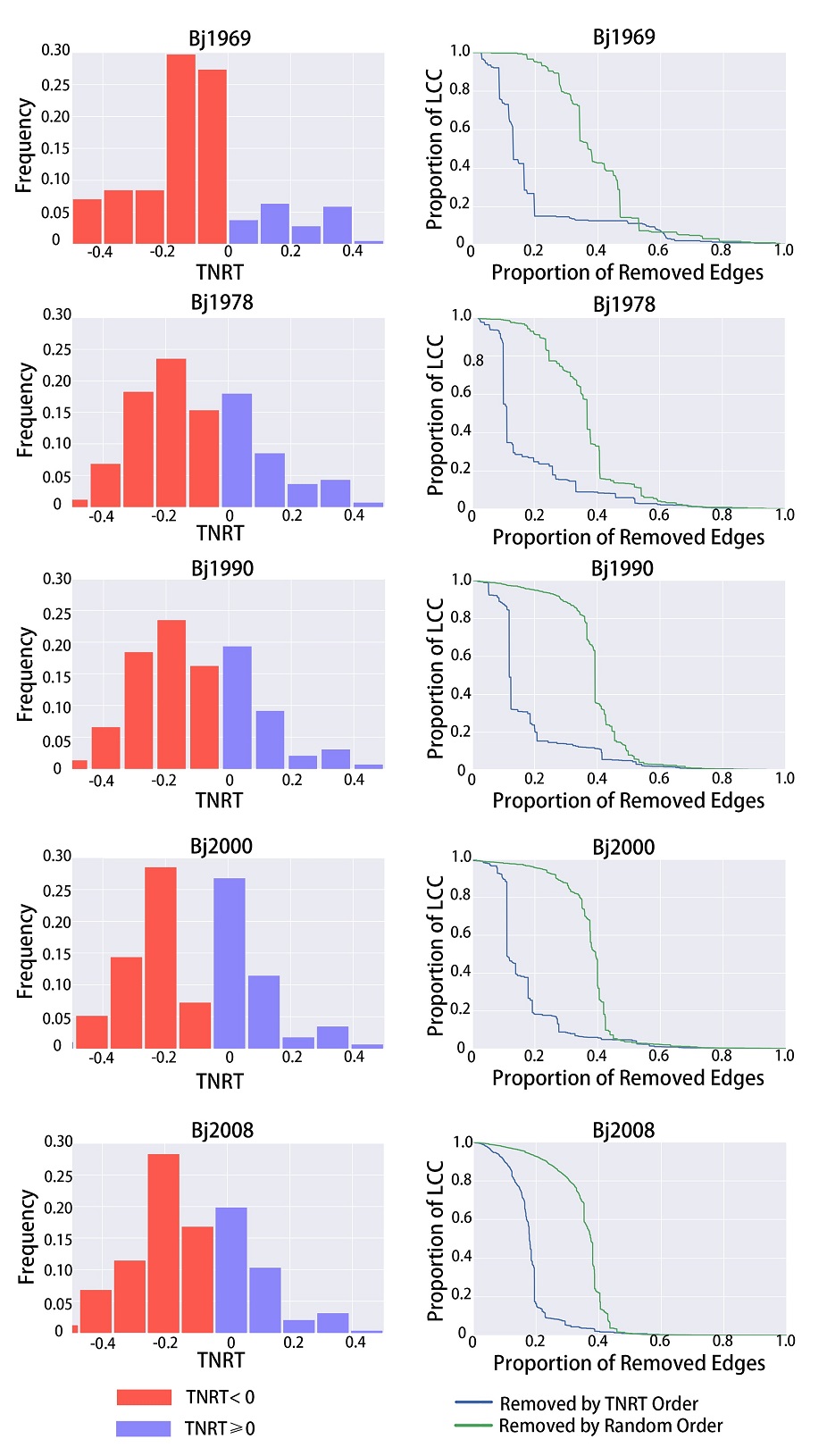}
\caption{TNRT distribution and TVR analysis in The evolution of road network}
\label{Figure 9}
\end{figure}

Then, following the previous attack strategy, we investigated the network’s vulnerabilities during different periods. When the simulation experiment is carried out according to the intentional attack mode, and the attack range accounts for 20\% of the whole network, only approximately 15\% of the network can operate normally. When the attack range exceeds 40\%, the network is paralyzed. In the same situation, when attacking randomly and the range accounts for 40\% of the whole network, all networks only have 40\% of normal operating capacity left, and when the attack range exceeds 50\%, the network is paralyzed. The experimental results show that the vulnerability of the road network is serious in every period, and with the evolution of the road network scale, the vulnerability of the road network does not change. These results should prompt road planners to rethink past models of construction and planning and once again emphasize the importance of road network vulnerability analysis.

In addition, we found the same rule as described in Section \ref{ST:4:2:1}. The damage to the network is greater during a purposeful attack than in a random attack, which are both reflected in the same range of attacks. For example, for the five road networks, when 20\% of the entire network is purposefully attacked, only approximately 15\% of the road network can be reserved for normal operation, while the network can retain approximately 80\% operating capacity during random attacks according to the Ricci curvature simulation. We emphasize that road planners need to protect road edges with a negative Ricci curvature lest they be interrupted and the whole network is affected.

\subsection{Different Vulnerability Methods Analysis} 
\label{ST:4:4}
The commonly used index in the vulnerability analysis of road networks is the betweenness centrality, which represents that the edges (nodes) that are often passed by the shortest path in the network. In this section, we will compare two different methods of the vulnerability indicators, and results are shown in \textbf{\textit{Figure}} \ref{Figure 10}.

The results of two intentional attack methods and a random attack method are shown in \textbf{\textit{Figure}} \ref{Figure 10}. The blue line is based on the Ricci curvature, the red line is based on the betweenness centrality and the green line represents a random attack. All urban networks show that an intentional attack is more harmful to the performance of the whole network than a random attack. Moreover, an attack based on Ricci curvature is more harmful to the network’s performance than one based on the centroid model. For example, with the Bj2008 road network, compared with the simulation attack based on the betweenness centrality, the attack based on the Ricci curvature causes a more rapid decline in network performance. When the attack range is approximately 20\%, only approximately 15\% of the networks can be kept operating normally, while the betweenness centrality can retain nearly 80\% operating capacity. Other cities show similar results, indicating that the Ricci curvature is a better indicator of the vulnerability for the critical structure of a network.
 
\begin{figure}
\label{Figure 10}
\centering\includegraphics[width=0.9\linewidth]{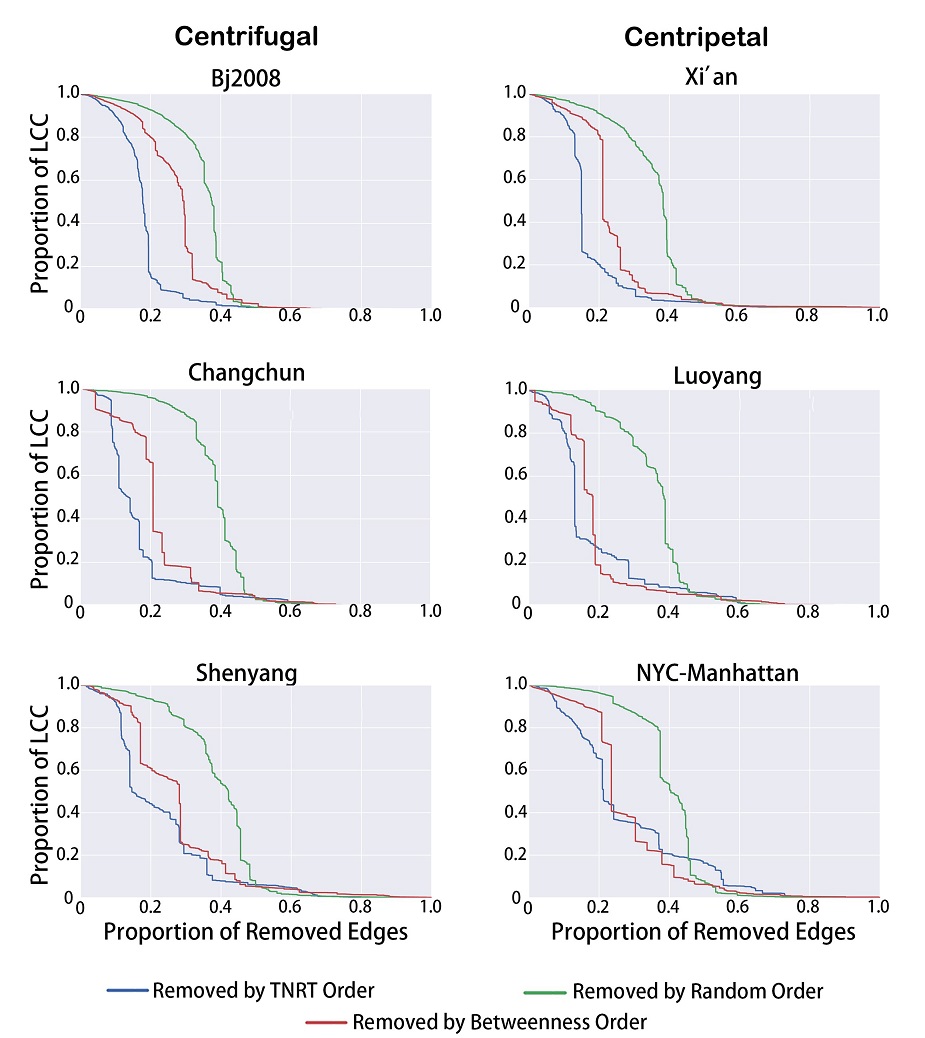}
\caption{Different vulnerability indicators}
\label{Figure 10}
\end{figure}

In addition, we tested the correlation between the Ricci curvature and the betweenness centrality of roads in all cities. \textbf{\textit{Figure}} \ref{Figure 11} shows a weak correlation between Ricci curvature values of a road and betweenness centrality values, and the correlation coefficients are almost below 0.1. This result was the same in all cities, and it further proves that the Ricci curvature for the network measurement is different from the betweenness centrality and can be used as a complement to topology measurements based on the edges of the complex network.

\begin{figure}
\label{Figure 11}
\centering\includegraphics[width=0.9\linewidth]{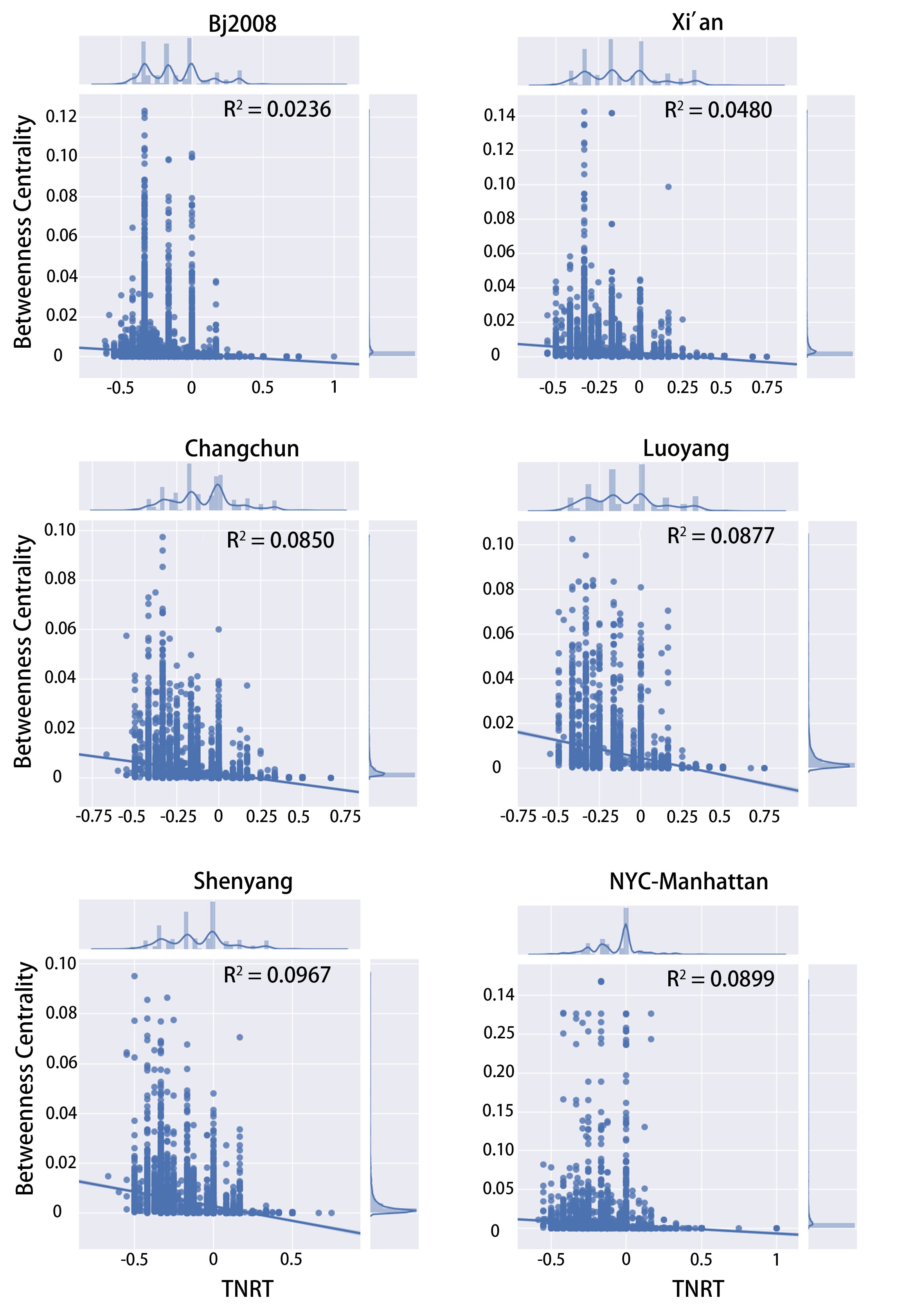}
\caption{Correlation of different indicators}
\label{Figure 11}
\end{figure}

Finally, we visualized the high-risk vulnerable roads discovered by the two methods as shown in \textbf{\textit{Figure}} \ref{Figure 12}. We find that the roads discovered based on our model in this paper are more localized, directly denoting the phenomenon that the topology transmission in the neighborhood around the roads does not match; conversely, the method based on the betweenness centrality shows a more global result. It focuses on expressing the concept of "transit", and the roads discovered are the sections that have been passed most frequently. This means that the two approaches demonstrate different aspects of vulnerability.

\begin{figure}
\label{Figure 12}
\centering\includegraphics[width=0.9\linewidth]{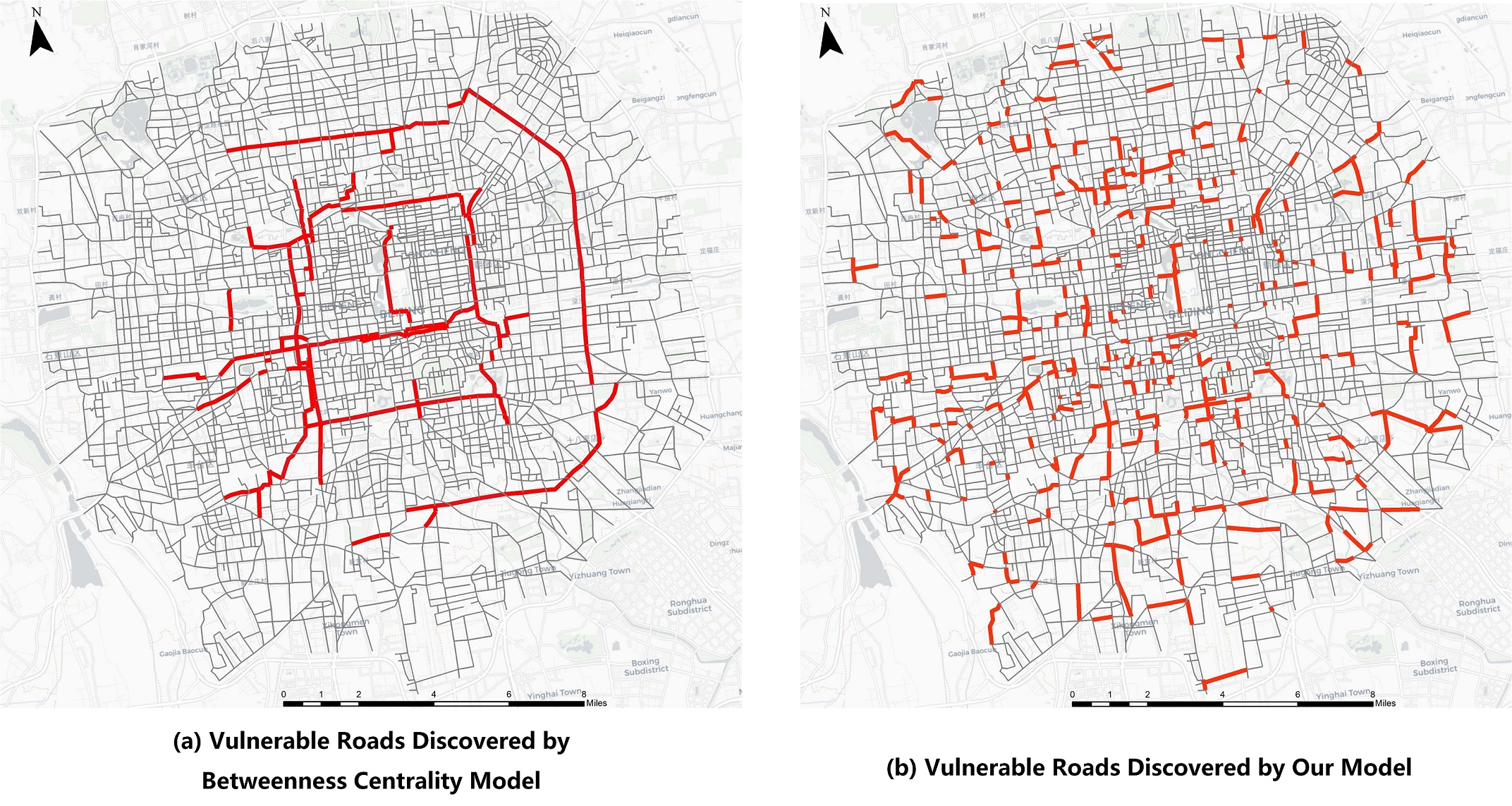}
\caption{Visualization of different indicators}
\label{Figure 12}
\end{figure}

\section{Discussion}
\label{S:5}
Vulnerability measurement is an important investigation factor in the public transport field. Most of the existing methods rely on single node information and therefore neglect the relationship between nodes. In this paper, we measured the topology transmission of the road network based on the edge-based Ricci curvature method. We then used these results to measure the topological vulnerability of the road network. 

The topology transmission based on underlying topology information determines the vulnerability. We calculated the road network topology transmission information among 6 different cities. We found that the negative curvature of the six cities is prominent, motivating us to review the meaning of negative curvature. 

By simulating the attack experiment, we note that a negative curvature implies vulnerability. Almost all networks show a greater vulnerability in a curvature-based attack, and the networks with centripetal components are even more obvious targets.

In a survey of vulnerability evolution in Beijing, we found that each period had a relatively high vulnerability with no significant changes. We suggest that road planners re-examine the road network’s planning and construction schemes and pay attention to the road network vulnerability.

Finally, we compare different measures of vulnerability; one is the betweenness centrality and the other is our model. We show that our model can better grasp the intrinsic and localized elements of the road network. At the same time, we point out that the two methods do not have a strong correlation.

\section{Conclusion and Future Works}

This paper constructs a measurement model of road network topology vulnerability based on the Ricci curvature of optimal transmission theory. The vulnerabilities of different cities and the evolution of vulnerability is also explored. By comparing our model with others, we display its advantages. The limitations, significant findings, and possible future improvements for this study are discussed below in detail.

\textbf{Data limitations}. First, we have obtained various forms of road network data to the best of our ability to fully investigate road network vulnerabilities. Although these are the best data currently available, the data set is not perfect. For example, the temporal resolution is low and some branches are lost in some cities. In addition, our research data set lacks some key information about roads, such as road width, road length, speed limit, and road level which makes us unable to set the weights of the edges effectively. When the weight is given, the measurement of the road is more consistent with reality, and the description of the vulnerability is more complete.

\textbf{Operational efficiency problem}. In this paper, the computational efficiency of the model is not analyzed in detail, and there may be potential operational efficiency problems. Later we may improve this model in terms of operational efficiency, especially regarding the method of calculating the Wasserstein Distance, such as Cuturi's approximate solution\cite{cuturi2013sinkhorn} and Wang’s alternative method\cite{wang2014bregman}.

\textbf{Demand model}. In this paper, the OD distribution and the relationship between road network vulnerability and demand distribution are not considered while building our model. The high-level performance of the road network is essentially the balance between supply and demand, and the nature of an urban road network can be described more completely by examining supply and demand at the same time.

\textbf{Network evolution}. A static network is adopted in this study. In the process of network evolution analysis, the time states are not rich enough, so a more detailed network evolution process will be put forward in future works. The evolution process of the network needs to reveal how the global network topology changes, such as the impacts of changes in the proportion of the entire network for centrifugal and centripetal components on the nature of the network. However, it is also necessary to reveal the changes over time in the road network. For example, with a city’s expansion, some roads that used to be branches may turn into main roads. During this process, the self-organizing characteristics, dynamics and the high order performances of the network are not clear and must be improved in the future.

\section{Acknowledgments}
This work was supported by the National Science Foundation of China (41571397, 41871364, 41830645, 41871302 and 41871276), by Natural Science Foundation of Hunan Province (2016JJ3144 and 2016JJ2006) and by the National Key Research and Development Program of China (No. 2016YFB0502601).



\section{References}


\bibliographystyle{model1-num-names}
\bibliography{ricci.bib}






\end{document}